\documentclass{aa}  

\usepackage{color}
\usepackage{bm}
\usepackage{amsmath,amssymb}

\begin{document}

\newcommand{\red}[1]{{\color{red}{\textit{#1}}}}
\newcommand{\blue}[1]{{\color{blue}{\textbf{#1}}}}

\title{GREGOR: Optics Redesign and Updates from 2018-2020}

\author{Lucia Kleint\inst{1} \and Thomas Berkefeld\inst{1} \and Miguel Esteves\inst{1} \and Thomas Sonner\inst{1} \and Reiner Volkmer\inst{1} \and Karin Gerber\inst{1} \and Felix Kr\"amer\inst{1} \and  Olivier Grassin\inst{1} \and Svetlana Berdyugina\inst{1}}
\institute{Leibniz-Institut f\"ur Sonnenphysik (KIS), Sch\"oneckstrasse 6, 79104 Freiburg, Germany}

  \date{Received April 20, 2020; accepted June 13, 2020}

\abstract
{The GREGOR telescope was inaugurated in 2012. In 2018, we started a complete upgrade, involving optics, alignment, instrumentation, mechanical upgrades for vibration reduction, updated control systems, and building enhancements and, in addition, adapted management and policies. This paper describes all major updates performed during this time.  Since 2012, all powered mirrors except for M1 were exchanged. Starting from 2020, GREGOR observes with diffraction-limited performance and a new optics and instrument layout.}


\maketitle

\section{Introduction}

Solar telescopes have always strived to evolve in diameter and thus spatial resolution \citep[e.g. review by][]{kleintgandorfer2017}. Before the year 2000, their diameters have remained below 1 m, with the exception of the Mc Math Pierce telescope at Kitt Peak, which however mostly observed in the infrared  \citep{pennlrsp2014}. A new generation of telescopes started in the 21st century with the Swedish Solar Telescope (SST) in 2002 with a clear aperture of 0.98 m, which was optimized for a very high image quality and routinely delivers impressive high-resolution solar images, especially also at wavelength in the blue \citep{scharmersst2003, scharmeretal2013}. It was followed in 2009 by the Goode Solar Telescope \citep[GST, ][]{goodeetal2010}, with a 1.6 m clear aperture, which for example has obtained the highest resolution flare images to date \citep{jingetal2016}. 

GREGOR, Europe's largest solar telescope, became operational a few years later.
Its 1.5 m diameter with an optical footprint of 1.44 m allows us to resolve structures on the Sun as small as 50 km at 400 nm. The GREGOR project started with its proposal in the year 2000 {\citep{gregor2001} and carried out a} science verification phase from 2012 to 2013. The state of GREGOR at that time was published in a series of articles in Astronomische Nachrichten Vol. 333, No. 9, in particular the GREGOR overview by \citet{schmidt2012}. {GREGOR was designed to explore solar features at smaller scales than other telescopes at that time. Its theoretical spatial resolution surpasses the SST and is similar to the GST and all three telescopes have significantly improved their image quality with state-of-the-art adaptive optics (AO) systems \citep{bbsomcao2016, berkefeld2018, scharmeretal2019}. Their designs differ though, with GREGOR focusing on high-precision polarimetry, which has enabled investigations of polarization signals as small as $10^{-4} I_c$ to detect spatial variations of turbulent magnetic fields \citep{biandaetal2018,dharaetal2019}. Another advantage of GREGOR is its potential for polarimetric night observations, which has been used to study the polarization of planets and thus their atmospheres \citep{gisleretal2016}. A past drawback of GREGOR was that its image quality did not reach the theoretical limit, partly because a risk was taken with untested technologies, such as silicon carbide mirrors, which could not be polished well enough and partly because of design issues. These issues have recently been solved by replacing all silicon carbide mirrors with mirrors made of Zerodur, which can be polished to the required quality, and by redesigning the AO relay optics, and GREGOR now operates at its diffraction limit. The goal of this paper is to summarize recent upgrades and enhancements that were carried out from 2018-2020. We will only briefly summarize GREGOR's general properties and we refer the reader to the article series from 2012 for more details.

 \begin{figure}[!htbp] 
  \centering 
   \includegraphics[width=0.41\textwidth]{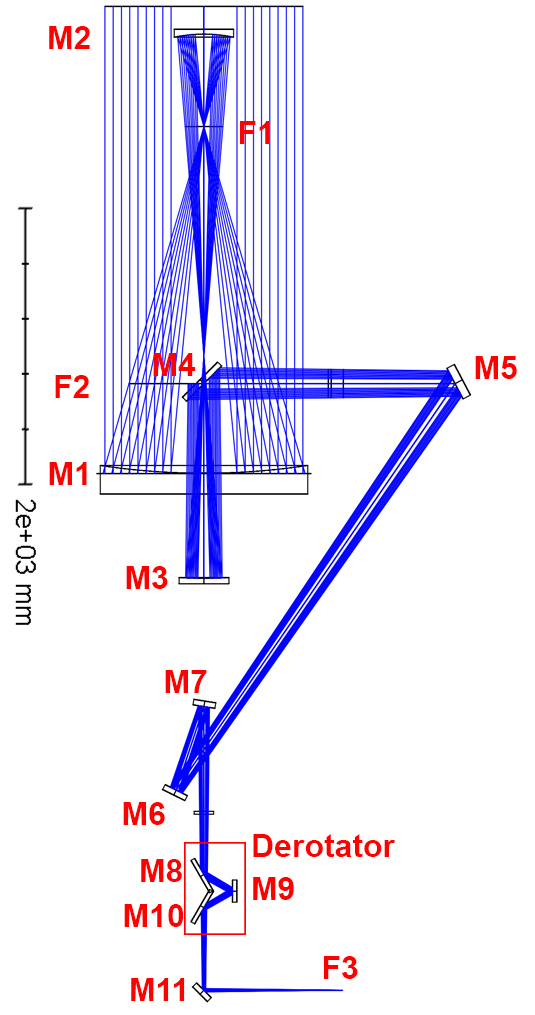}
   \caption{GREGOR optical layout until F3}
        \label{gregor}
  \end{figure}
%

GREGOR obtained its name by being a Gregory system with three imaging mirrors (M1, M2, M3) whose properties are summarized in Table~\ref{tabmirror}. More than 99\% of the sunlight is reflected away at the cooled primary field stop F1.  Only a beam with circular diameter of 150\arcsec\ passes through its central hole to M2. The F1 field stop is recoated yearly, currently with an aluminum layer on top of a nickel layer. The mirrors M4-M11 are flat mirrors with the purpose of directing the beam into the optics lab one story below the telescope level. M8-M10 are rotatable about the optical axis, thus acting as a derotator, which compensates for the solar image rotation induced by the alt-az mount of the telescope. A schematic drawing of GREGOR is shown in Fig.~\ref{gregor}.

\begin{table*}[bth]
\begin{tabular}{l | l l l}\label{tabmirror}
& M1 & M2 & M3\\
\hline
focal length [mm] & 2506.35 & 519.40 & 1398.5 \\
optical footprint for 150\arcsec\ [cm] & 144 & 38.1 & 27.5  \\
f\# & 1.7 & 1.4 & 5.1 \\
curvature radius [mm] & 5012.70 & 1039.79 & 2797.00 \\
conic constant & -1 (parabolic) & -0.306 (elliptic) & -0.538 (elliptic) \\
\hline
\end{tabular}
\caption{Mirror properties for M1-M3.}
\end{table*}

   \begin{figure*}[tb] 
   \centering 
   \includegraphics[width=0.8\textwidth]{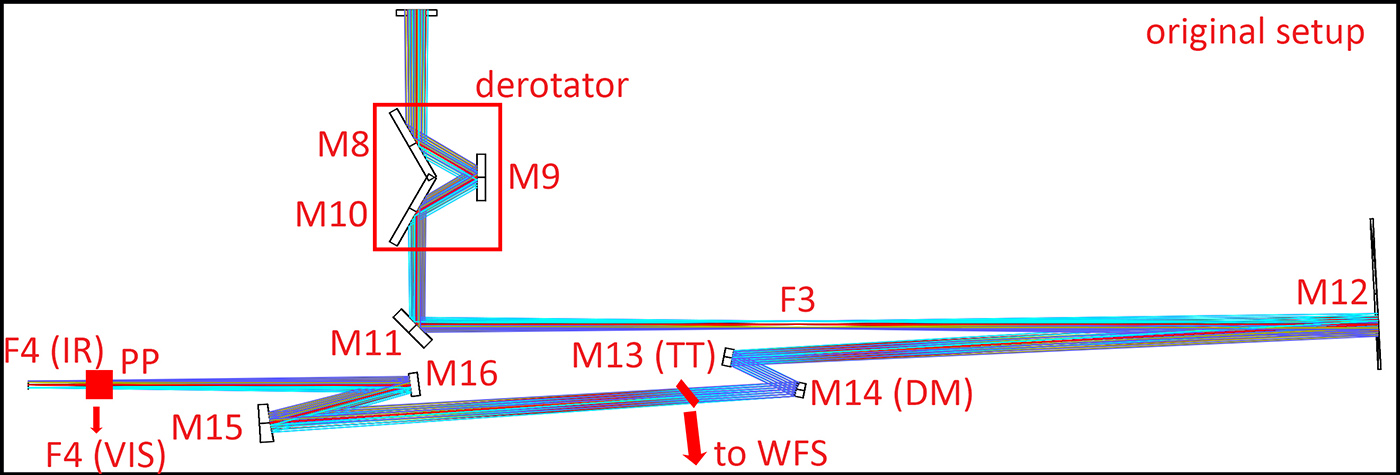}\vspace{0.5cm}
   \includegraphics[width=0.8\textwidth]{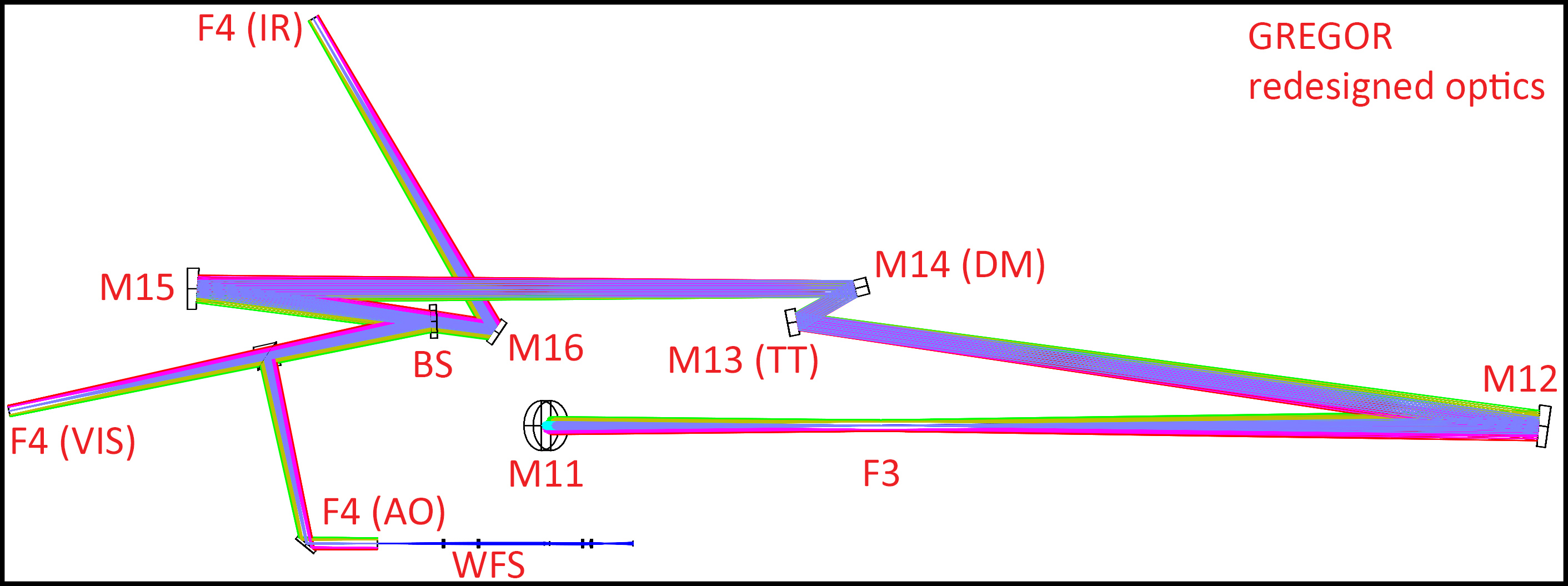}
   \caption{\textbf{Top image}: Original GREGOR setup in side view from the vacuum exit window to F4. The location of the wavefront sensor (WFS) is indicated with an arrow. The wavelength splitting was done with a pentaprism (PP) and the instrument focal planes (F4) are shown for IR and VIS.
   {\textbf{Bottom image}: New setup in top view. The VIS beam is reflected on the front side of the beamsplitter plate (BS). It is then split by a beamsplitter with a fraction of the light going into the AO. The IR beam passes BS and is then reflected on M16.  The WFS is located after all mirrors with power.}}
        \label{design}
  \end{figure*}

\section{Optics}\label{obs}

\subsection{Redesign of the optics lab}
The original optics lab layout was devised during GREGOR's design phase before 2008. It focused on the first light instruments GRIS, a spectropolarimeter based on a grating spectrograph \citep{colladosgris2012}, GFPI, a dual-etalon spectroscopic imager \citep{puschmann2012},  plus an associated broad-band imager, and BBI as a standalone imager \citep{bbi2012}. Guest instruments, such as ZIMPOL for high-precision spectropolarimetry \citep{zimpol1997}, or the GREGOR planet polarimeter \citep[GPP,][]{gisleretal2016} were also regularly operated. The adaptive optics (AO) was mounted on a vertical bench, which saves space, but may be disadvantageous in terms of vibrations and alignment. The beam is collimated for the AO after F3 via M12 and then reimaged via M15. The AO itself consists of a tip-tilt (M13) and a deformable mirror (DM, M14). The setup is shown in the top panel of Fig.~\ref{design}.
During the past two years, we realized that there were issues with beam stability, alignment, and the optical quality induced by the two biconic mirrors M12 and M15. These mirrors originally featured biconic shapes to minimize the astigmatism due to an oblique incidence on a focusing mirror. However, we noticed that they created field-dependent aberrations (mostly astigmatism and coma), prominent at the design angles. Unfortunately, M12 and M15 could not be aligned arbitrarily to minimize the aberrations, which however never fully vanished, because of a fixed focal plane and vignetting at other optical elements, plus their alignment tolerances were too strict. Additionally, there was a lack of space in the optics lab to develop and install new instrumentation. Therefore, we completely redesigned the optics after M11, including new off-axis parabolic mirrors to replace M12 and M15 and an improved instrument layout.

The new layout was devised based on the following criteria: 
\begin{itemize}
\setlength\itemsep{.1em}
\item More space for the science instruments and a future multi-conjugate adaptive optics (MCAO) via a different beam distribution while keeping all current capabilities.
\item Horizontal setup of the AO relay optics (for vibration reduction, stability, and easier
alignment)
\item (Total) angle at M12/M15 = 8$^\circ$ (the two mirrors compensate each other). Angles $ <8^\circ$
are not possible due to the beamsplitter plate between M15 and M16.
\item 1:1 imaging of the AO relay optics between F3 and F4. Exit pupil telecentric (at infinity).
\item Perfect image quality in F4 over a radius of 60" (=2 arcmin image diameter).
\item F3 at least 1000 mm after M11.
\item pupil size limited by the DM size.
\item A visible/infrared (VIS/IR) beamsplitter between M15 and M16 reflects the VIS, only IR passes through towards
GRIS. This improves the antireflection coatings both for the VIS and the IR elements. This beamsplitter is mounted on a rotating table and can be exchanged by e.g. a 50/50 beam splitter for special setups that require other wavelength distributions.
\item Small angles at mirrors for polarimetry.
\item DM-M15 parallel to M11-M12 (not strictly required, but simplifies the alignment).
\item F4 IR unchanged (1420 mm from M11). M16-F4 IR shall be on the same line as M11-F4 IR to keep the beam angles, such that the GRIS instrument does not require any changes.
\item The wavefront sensor (WFS) shall be located after all powered mirrors.
\end{itemize}

 \begin{figure*}[tb] 
  \centering 
   \includegraphics[width=0.45\textwidth]{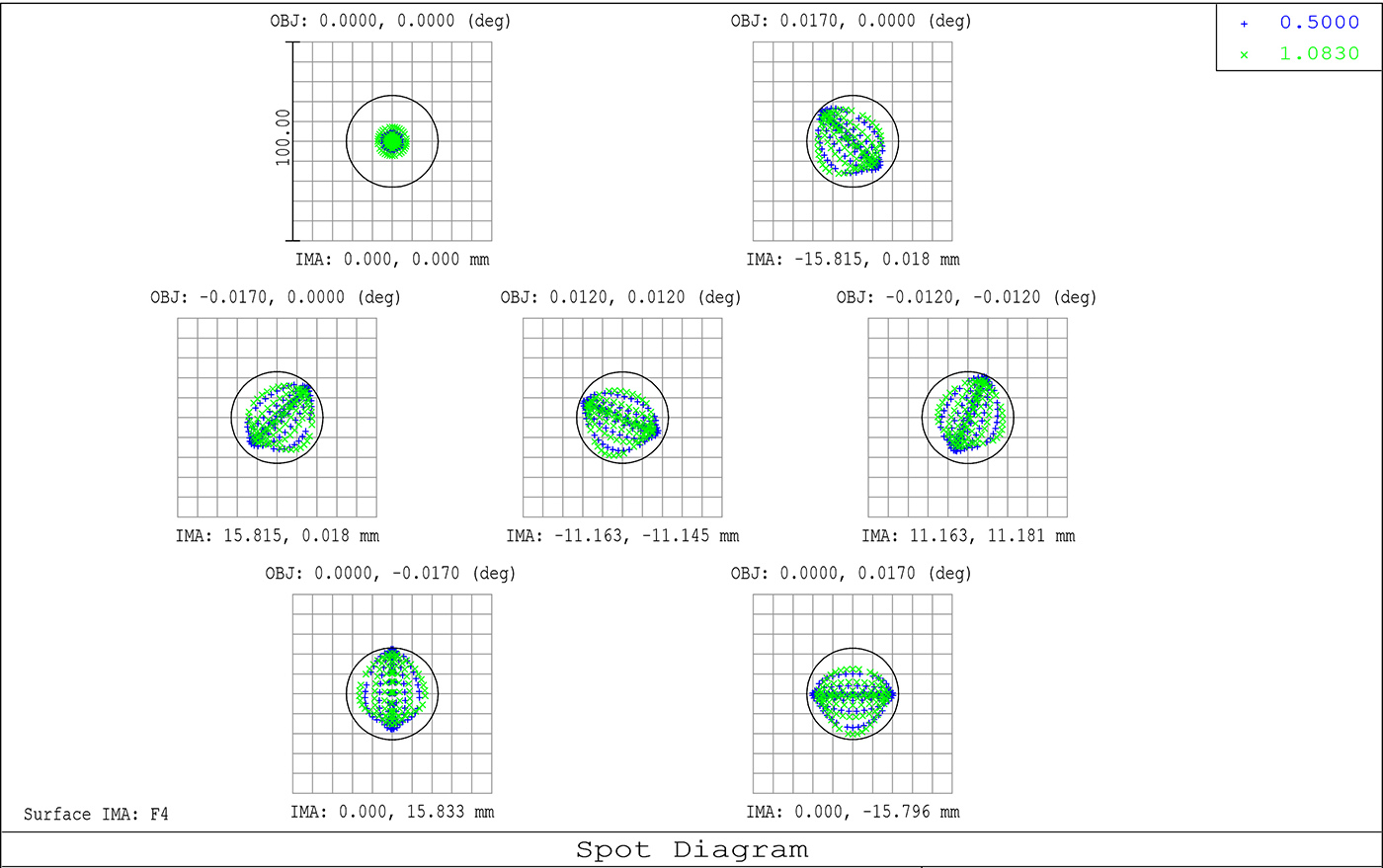}
   \includegraphics[width=0.45\textwidth]{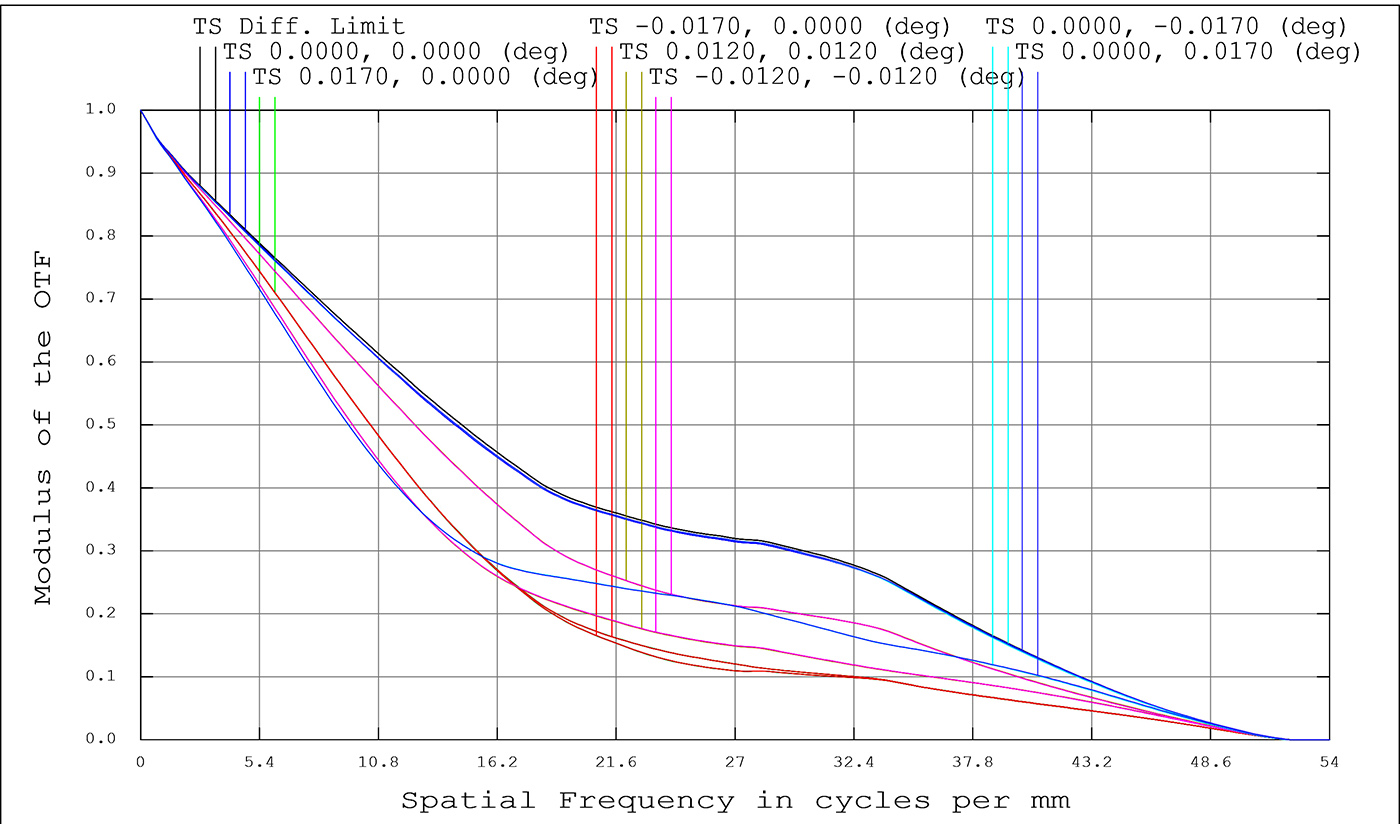}\vspace{0.1cm}
 
   \noindent\makebox[\linewidth]{\rule{\linewidth}{0.4pt}}
   \vspace{0.1cm}

   \includegraphics[width=0.45\textwidth]{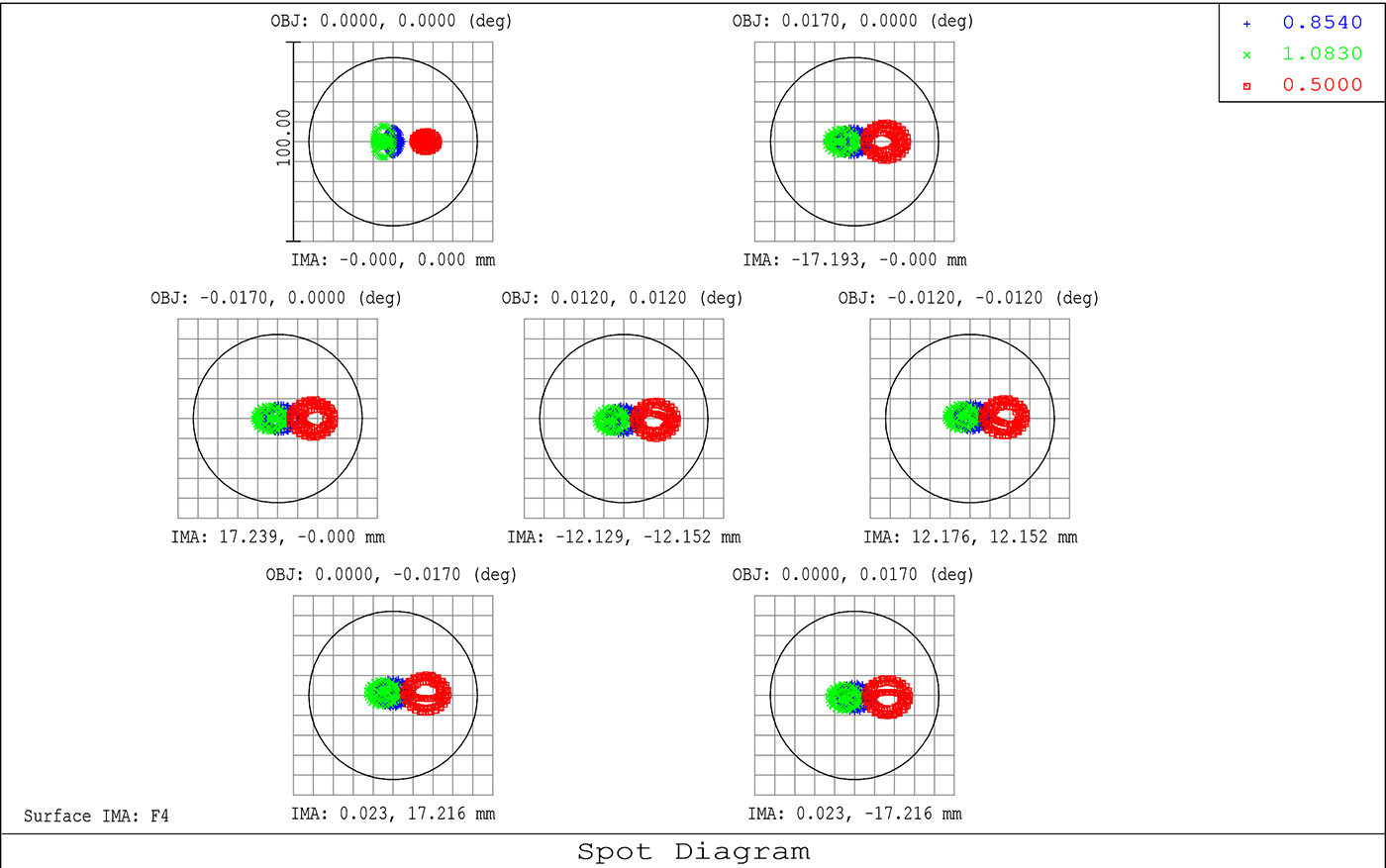}
   \includegraphics[width=0.45\textwidth]{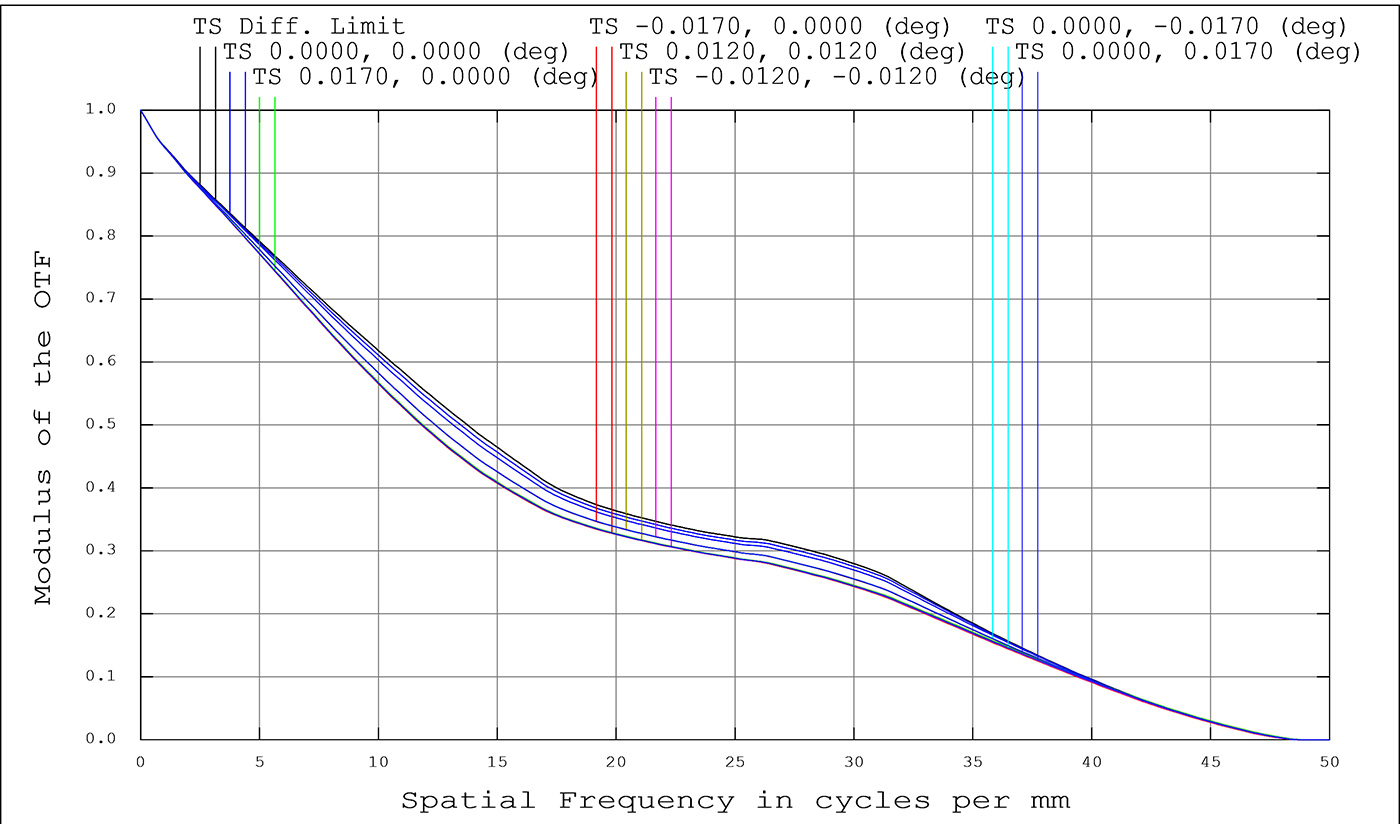}
   \includegraphics[width=0.45\textwidth]{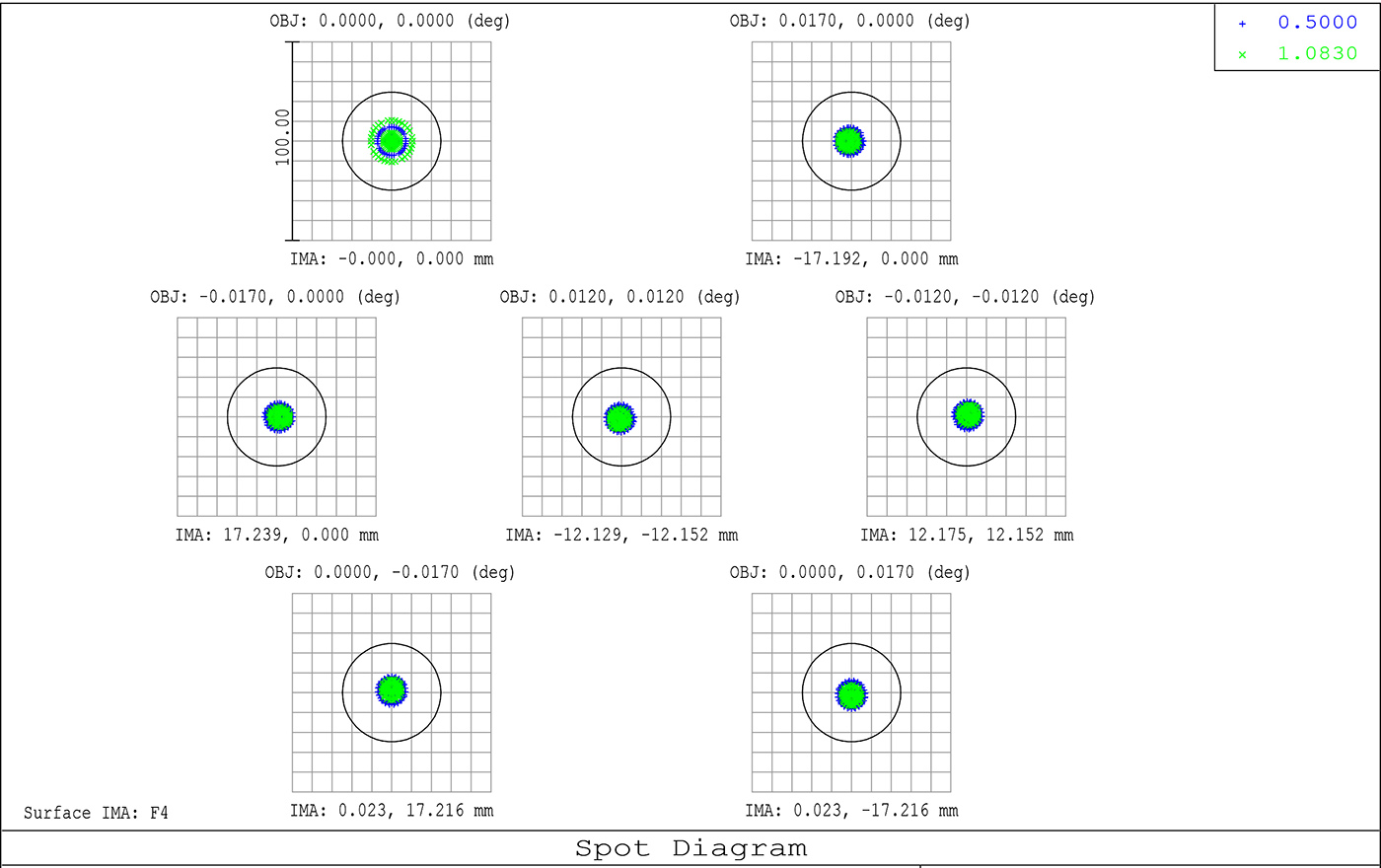}
   \includegraphics[width=0.45\textwidth]{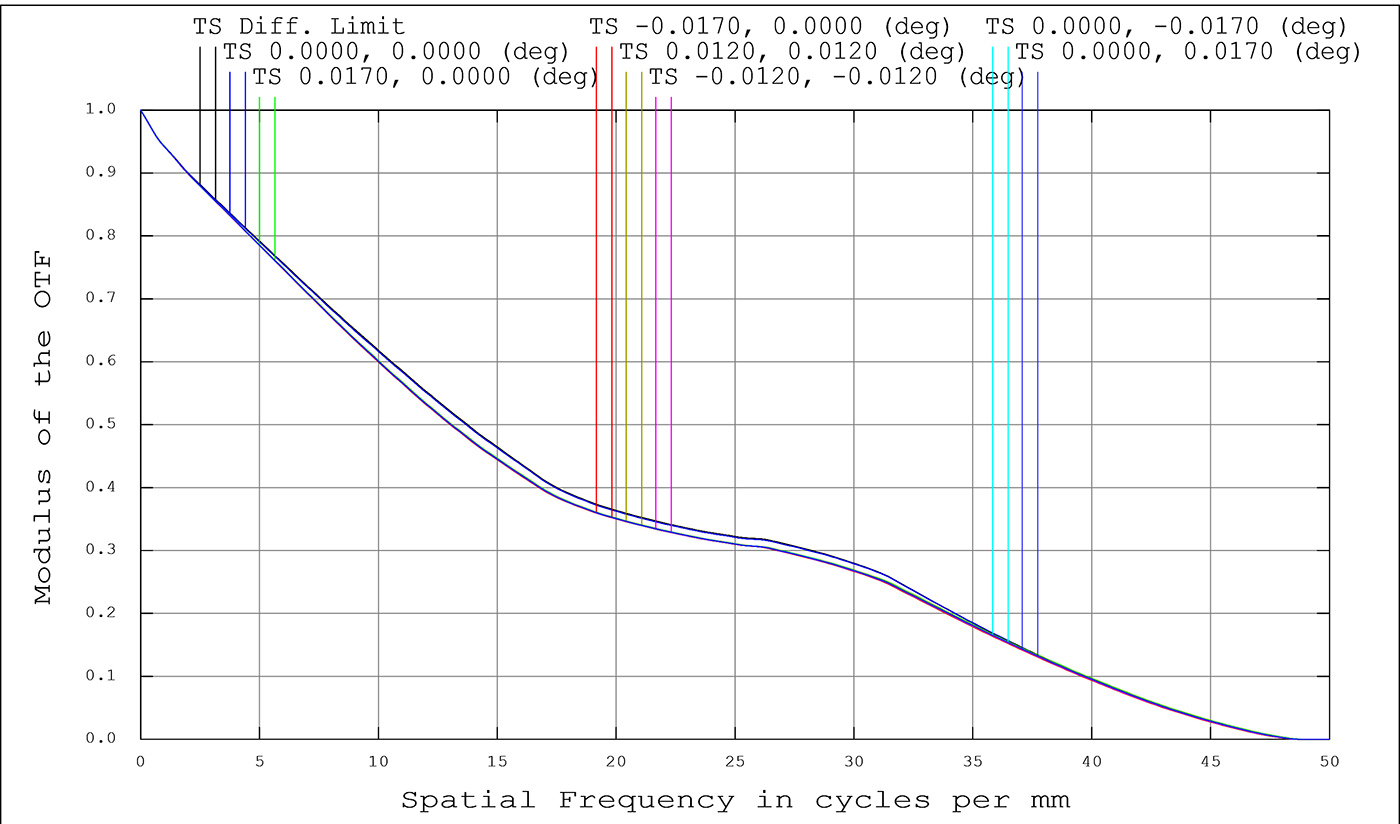}
   \caption{GREGOR spot diagrams for different wavelengths and MTF for 500 nm with max.\,FOV=120\arcsec. The circle indicates the Airy disk. Top row: original M12 and M15. Diffraction-limited performance is only achieved for the center of the FOV with a very fast drop in contrast off-center. Note that this simulation is done for ideal M12 and M15 mirrors and our suspicion is that their specs were not met, which would further worsen their properties. Middle row: New IR setup. The lateral shift is due to the beamsplitter plate and 500 nm (red) is displayed to show that even the ZIMPOL guest instrument setup, which uses VIS in the IR beam by using a custom beamsplitter plate, would still have diffraction-limited performance. Bottom row: New visible setup with basically no loss of contrast for the whole field of 120\arcsec. Note that all MTFs have good properties in the IR and therefore 500 nm is shown here.}
        \label{mtfpsf}
  \end{figure*}
%


 \begin{figure*}[tb] 
   \centering 
   \includegraphics[width=0.65\textwidth]{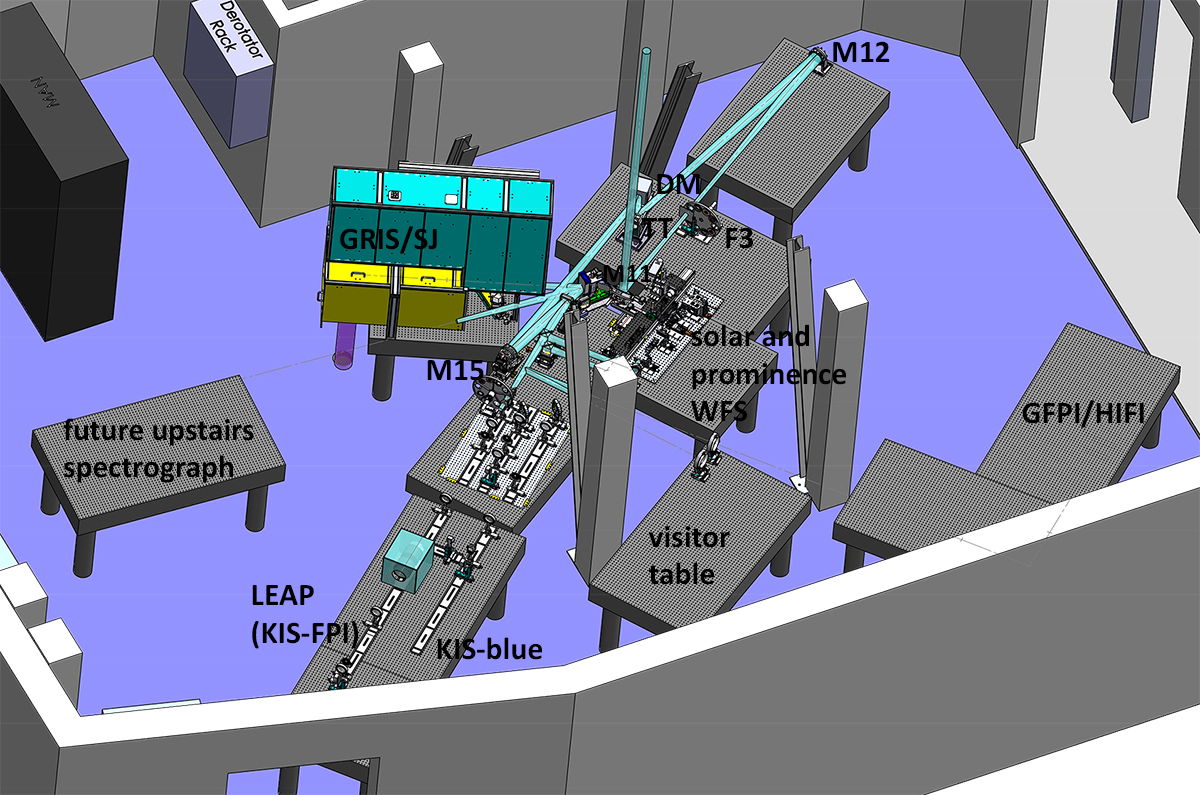} 
    \includegraphics[width=0.34\textwidth]{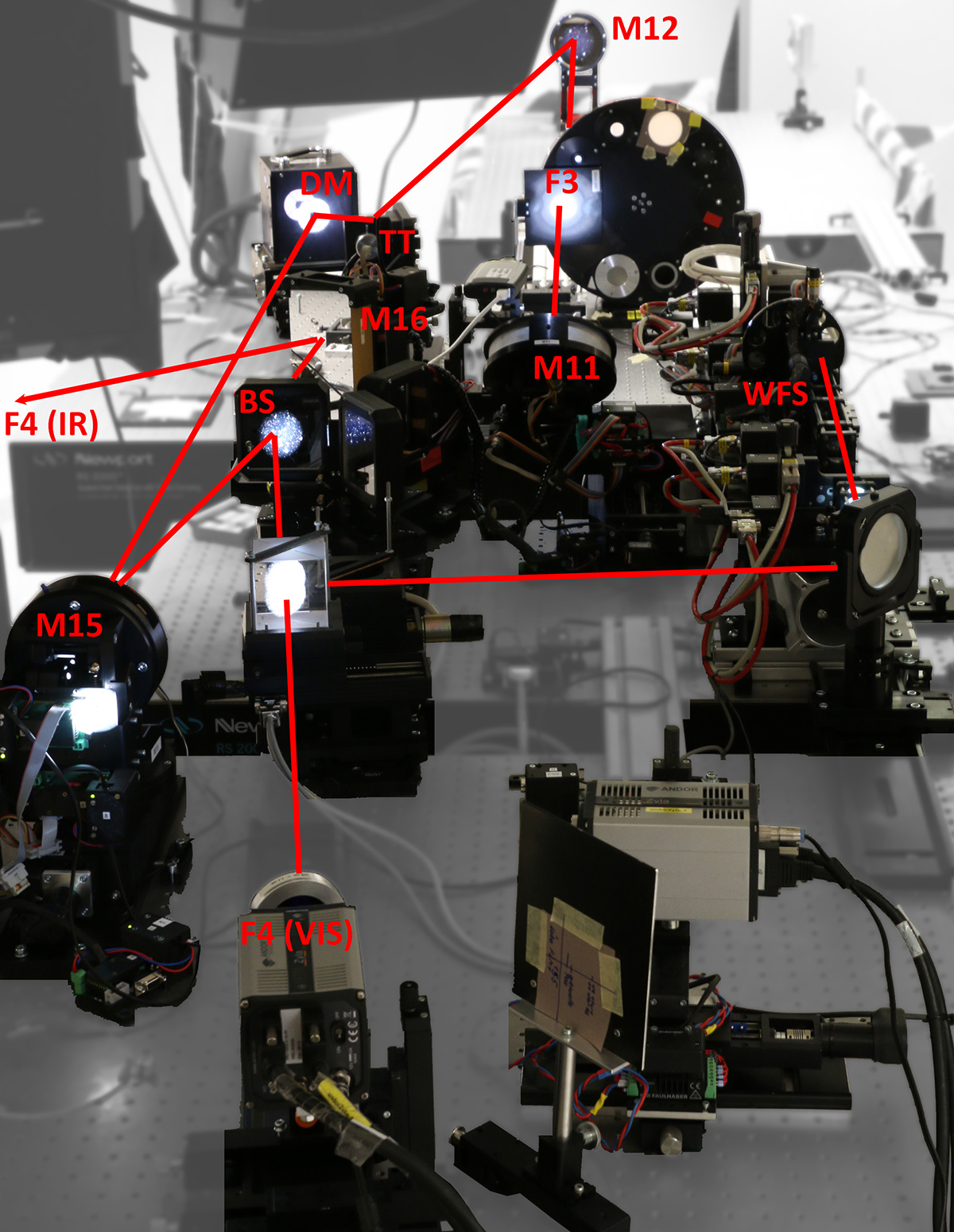}\vspace{0.5cm}
  \caption{Left: New instrument layout at GREGOR. GRIS (yellow) with the SJ system (green box above GRIS) are the only optics that remain in the same place, but all original instrument combinations remain. The AO is now horizontal and the new KIS FPI is drawn towards the window. The empty tables are for temporary visitor setups and for GFPI/HIFI. Right: Photo of the setup as installed in March 2020. The red line traces the beam.}
        \label{instr}
  \end{figure*}

%

 \begin{figure*}[tb] 
   \centering 
   \includegraphics[width=0.8\textwidth]{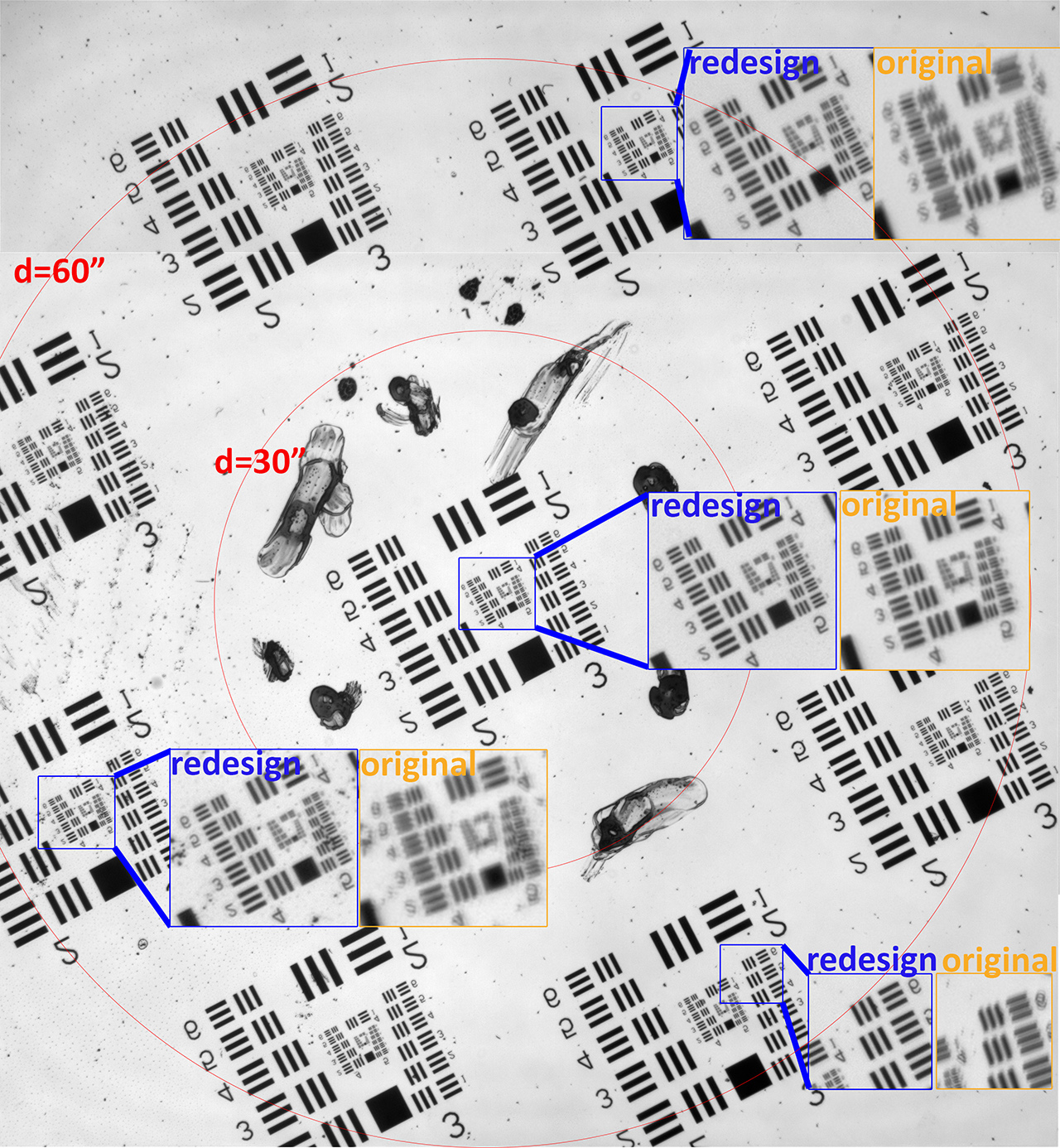} 
  \caption{Image of a test target obtained after the redesign. The blue insets show magnified sections, the orange insets demonstrate the comparison to the original optics setup, which suffered from astigmatism and coma outside of the image center. The redesigned optics are perfectly sharp across the whole field of view. The image scale is indicated with red circles, whose diameters are specified in the figure.}
        \label{target}
  \end{figure*}


The original and the new design are shown in Fig.~\ref{design}. The new off-axis parabolic (OAP) mirrors M12 and M15 are identical and compensate each other's aberrations. Their paraxial focal length is 1978 mm and the off-axis distance is 277 mm. They are made from Zerodur with a diameter of 125 mm, thickness of 20 mm, with a silver coating. Their measured surface error RMS is 6.9 and 5.2 nm for the two mirrors, and this quality was achieved with two ion beam figuring (IBF) polishing runs per mirror.

The original and new MTFs and spot diagrams are shown in Fig.~\ref{mtfpsf}. The MTF for different field angles improves with increasing wavelength, but obviously the spatial resolution decreases at higher wavelength. We therefore show the MTF at 500 nm to illustrate the loss of contrast issues of the original setup and the much better performance of OAPs. All plots assume a maximum field of view (FOV) of 120\arcsec.

Figure~\ref{instr} shows the new instrument layout and a photo of the setup in March 2020. GRIS and the slitjaw system remain. The new Fabry-Perot Interferometer (FPI), currently under development, will be in the straight beam and has minimal reflections. Parallel to it is its broadband channel for which we plan phase diversity, and a fast imager in the blue. An insertable mirror, which can be rotated, sends the light either to a visitor table, or to GFPI/HIFI, which is now located near the entrance to the optics lab. One table is saved for a future upstairs spectrograph, which can be fed by rotating M16. {About 75\% of the mechanical parts could be reused and were adapted for the new setup, which accelerated the construction and fabrication.} 

The new design has many advantages: By having a constant beam height above the optical tables, all alignment is simplified. The off-axis parabolas have much better alignment tolerances than the former biconic mirrors. The PSF remains near diffraction-limited in all field points, even for small deviations of properties (e.g. conic constant difference of 2\%, decenter of 5 mm, 1\% deviation in paraxial radius). The MTF remains nearly constant across the whole FOV, compared to a $\sim$50\% drop in the original setup. There is space for a future MCAO and a future upstairs spectrograph, while all current instrument capabilities remain. Additionally, all optical elements are accessible. Furthermore, all critical optical elements are motorized, so that their alignment is reproducible. During the redesign, the number of motorized elements was increased from 7 to 13 and the motorization control interfaces were updated for manual and remote control. The wavefront sensor is located after M15, thus able to correct for any static aberrations induced by the relay optics. The wavelength splitting is performed using a dichroic beamsplitter plate and no longer a pentaprism, which is an advantage for polarimetry (smaller angles). By using different beamsplitter plates, it is possible to change the wavelength cutoff for the visible beam, currently either to 650 nm or 900 nm. The location of the beamsplitter plate in the beam simplifies the IR and visible coatings because now only the rear side of the beamsplitter plate requires a broad antireflection coating, while instruments receive limited wavelength ranges and therefore can use mostly standard antireflection coatings, which are easier to obtain.

The new optical setup was installed in March 2020. The optical elements were aligned with a laser, which was tracing the beam backwards from F4. The laser direction was aligned in azimuth from the original F4 GRIS location to a plummet hanging from the exit window, which defined the vertical optical axis of the telescope. During the backwards tracing, the beam angles were measured for M15, the DM and M12 and the mirrors shifted and tilted if necessary to match the design values. The laser was reflected upwards into the vertical optical axis by M11, where it coincided with the plummet line. The wavefront sensor was only adapted to its change of location, but its properties remain.  We then calibrated the AO by applying Hadamard shapes to the DM and measuring the response of the wavefront sensor. This gave us the response of the 256 actuators. A pinhole in the AO F4 was used to measure the subaperture reference positions for the perfect wavefront. Then we locked the AO on a USAF target in F3, which was immediately sharp over the whole field of view in the science camera in F4 VIS. The improvement compared to the old relay optics was immediately noticeable and can be attributed to the pair of off-axis parabolic mirrors with a small and equal reflection angle.

Figure~\ref{target} shows a test target recorded on March 21, 2020. It was recorded directly in the visible F4, with an exposure time of 3 ms at $393.55 \pm 1.1$ nm. No AO mode offsets were applied or necessary. The pixel scale is 0\farcs023 pixel$^{-1}$. The figure is a composite of two images: the camera was shifted in height to record all subtargets. This figure demonstrates that the redesign was successful because all targets are sharp across the whole field of view and the correct element (5th group, 5th element) is resolved in all subtargets, some of which are shown magnified in the blue boxes. No astigmatism or coma is visible. 
For comparison, the orange boxes show a test target, taken with the original optics at 396 nm on July 19, 2019. It is clearly visible that previously, locations outside of the image center showed strong astigmatism (``shadows''), consistent with the MTF shown in Fig.~\ref{mtfpsf}.

 \begin{figure*}[tb] 
  \centering 
   \includegraphics[width=.9\textwidth]{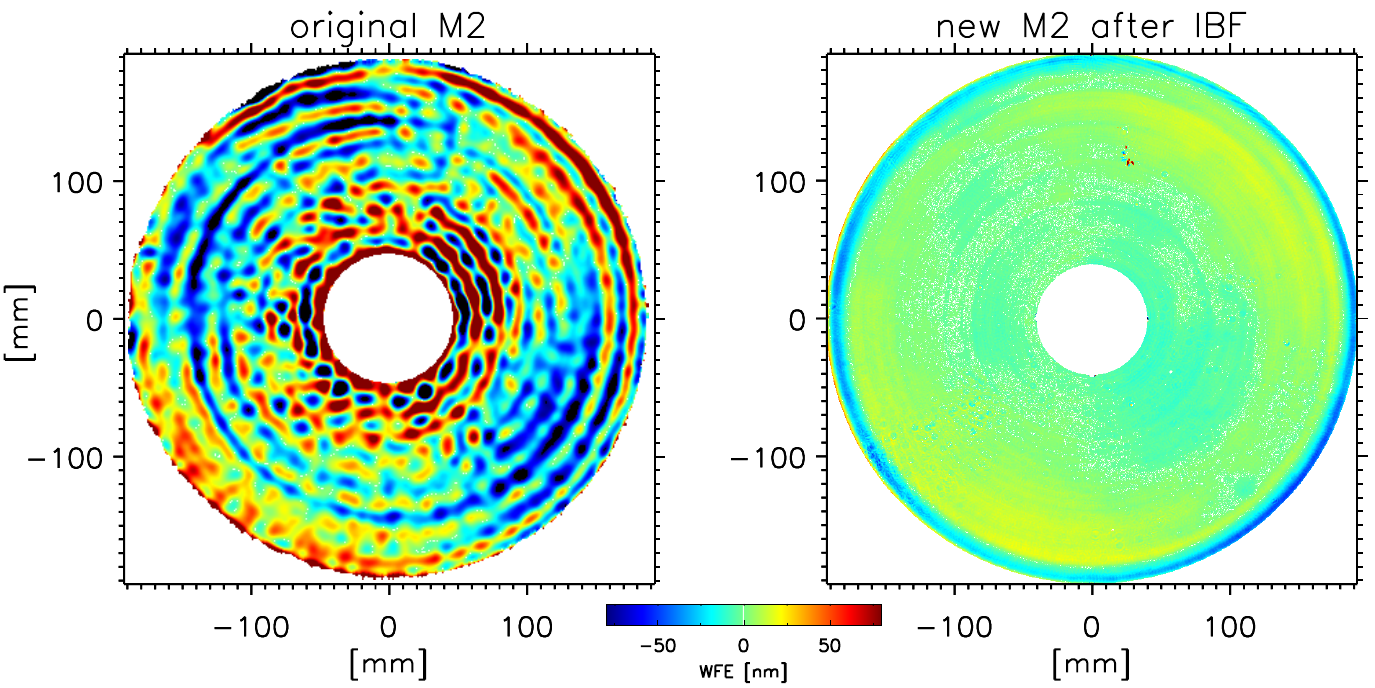}
   \caption{Comparison of the interferograms of original and new M2. The display range is $\pm$80 nm for both, showing wavefront errors, which are a factor of 2 larger than surface errors. Some dust and finger prints explain the small imperfections on the interferogram of the new M2. The WFE rms is 39 nm and 8 nm for original and new M2, respectively.}
        \label{interferogram}
  \end{figure*}

 \begin{figure}[tb] 
   \centering 
   \includegraphics[width=.5\textwidth]{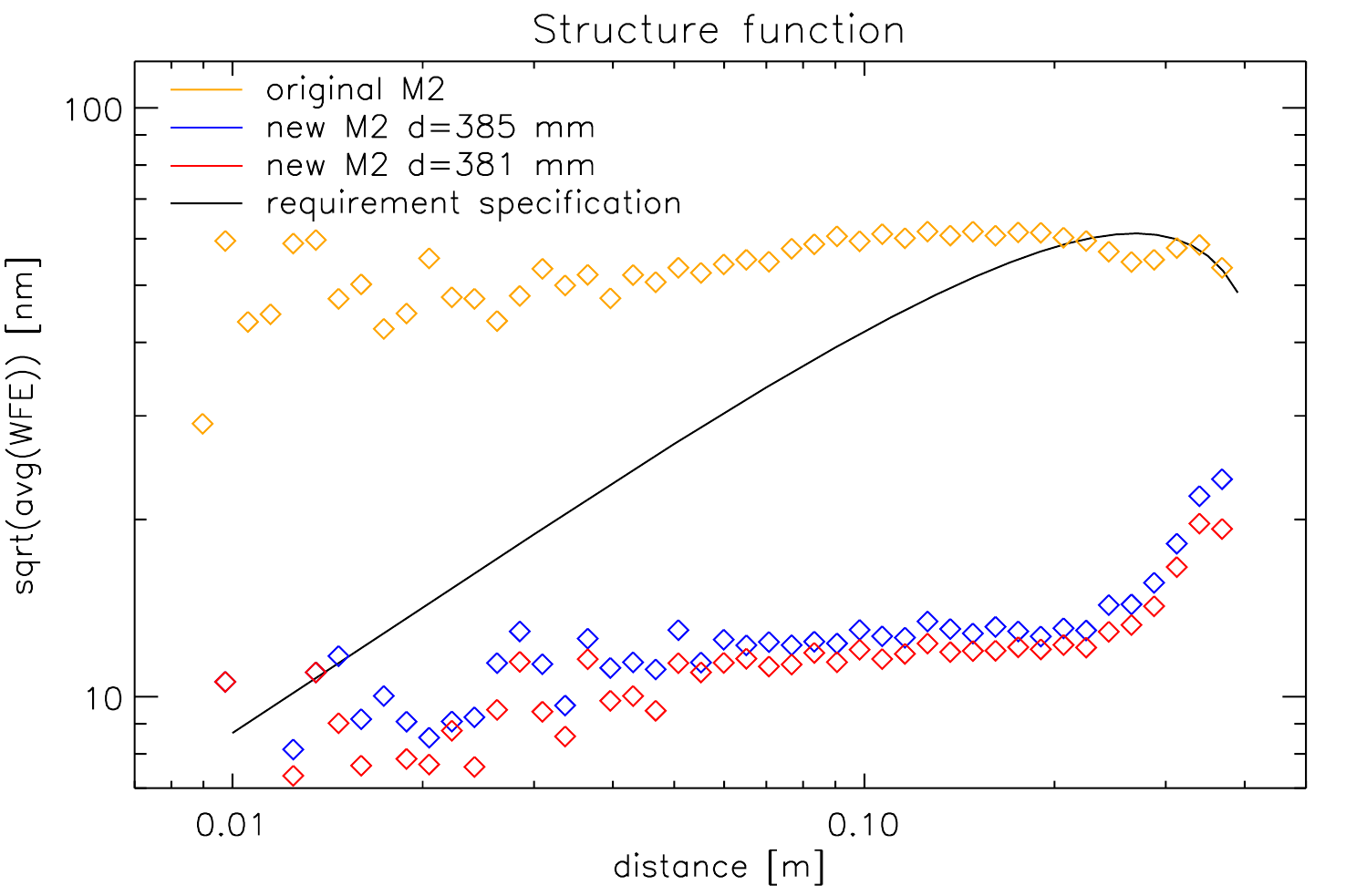}
   \caption{Structure function. 10$^6$ randomly selected points were used to calculate differences in the WFE according to Eq.~\ref{eq1}. The new M2 is significantly better than the requirements. The larger scatter near smaller spatial scales is due to averaging fewer points when binning. The current optical footprint on M2 is 381 mm.}
        \label{strucfunc}
  \end{figure}

\subsection{M2}
The original secondary mirror (M2) was fabricated from silicone carbide (Cesic) a material of high stiffness, whose thermal properties are favorable to absorb the 1.3 W from the incoming sunlight after the primary field stop when assuming that each mirror absorbs 10\% of its incoming light. Unfortunately, the Cesic surface could not be polished well enough and in the end of the polishing the original M2 had an rms wavefront error (WFE) of $\sim$40 nm, mostly in the shape of concentric rings as shown in Fig.~\ref{interferogram}. These rings were of the same scale size as the AO subapertures, which means that the AO could not ``see'' them and therefore was not able to correct them. 
Such mid-frequency errors lead to a severe reduction in contrast both in the WFS and in the science camera. The low WFS contrast then leads to a low signal to noise ratio of the wavefront reconstruction and a reduced performance of the AO, especially when locking on granulation. 
Furthermore, the unsensed mid-frequency errors lead to spatial aliasing, which means that insufficient spatial points are available to determine the wavefront, and result in non-optimal wavefront reconstruction and correction / increased wavefront error, which is why we could not fully exploit the maximum number of corrected modes in the past.

It was therefore decided in 2014 to replace M2 with one made of Zerodur. A call for tender was issued, this time with a much stricter specification on the allowed WFE. The WFE $\Phi$ was specified in terms of a structure function $D_{\Phi}(r)$, which describes the error at different spatial scales:
\begin{equation}\label{eq1}
D_{\Phi}(r) = avg(\left[ \Phi(x+r) - \Phi(r)\right]^2)
\end{equation}
where r denotes different separations of (randomly) selected points. The outer diameter was specified to 430 mm. Because M1 is stopped down to 1.44 m due to its polishing quality, the optical footprint on M2 is 381 mm, but up to 385 mm are of sufficient polishing quality.

There were some issues during fabrication and the specs could not be met with conventional polishing methods. However, after using IBF, the mirror's surface is now among the best at GREGOR. The new M2 was installed in July 2018 and the improvements in contrast after its installation at GREGOR were noticeable even before carrying out any acceptance testing. They are difficult to quantify, but we estimate an improvement of the contrast of a factor 1.5-2 in the WFS. Since then, the adaptive optics easily locks on granulation, even during bad seeing. Figure~\ref{interferogram} shows the interferograms of the original and the new M2. To display both on the same scale ($\pm$80 nm WFE), data from the original M2 had to be clipped. Figure~\ref{strucfunc} shows the structure functions of original and new M2 and the specifications from the call for tender, which were clearly exceeded. This was the last Cesic mirror to be replaced and GREGOR now contains only Zerodur mirrors.

\subsection{Derotator}
The derotator consists of M8-M10 and was installed in 2016. M8 and M10 are tilted by 30$^\circ$, leading to incident beam angles of 60$^\circ$, which is not optimal for polarimetry and requires careful calibration {\citep{hofmannetal2012}}. By rotating about the optical axis, the derotator counteracts the image rotation introduced by the alt-az telescope mount. The image rotation is the sum of parallactic angle + azimuth - zenith distance - solar p0 angle + an offset for the desired orientation. The required derotator rotation is half of the image rotation. The image rotation rate is highest in June when the Sun nearly reaches zenith and exceeds 5$^\circ$ per minute then, while it generally remains below 0.5$^\circ$ per minute before 11 UT during nearly the whole year.

The original developer GUI was used by observers to control the derotator for several years. In 2019 we enhanced the derotator control in functionality and design. The derotator can now be controlled from the GREGOR GUI and we calibrated its orientation, such that observers can simply insert the desired solar angle (e.g. 0 for north-south) for the spectrograph slit orientation and all calculations are performed internally.

\subsection{Alignment}
GREGOR has very tight alignment tolerances due to its fast beam. It is therefore nearly impossible to align the telescope visually (using lasers) and past experience has shown larger aberrations after each recoating. It is possible that each M1 removal and remounting has modified its position because even a 0.001 degree tilt is noticeable in the aberrations measured by the AO, but not visually. In the past, M2 was shifted and tilted to counteract these aberrations, but after the fourth recoating in 2017, the required shifts reached several mm.

This issue was solved in two ways: 1) We devised a new method to align M1 by using the AO and successfully carried out the alignment in May 2019. The method involves using the force actuators on the back of M1 and creating a ``response matrix'' of each actuator. It can then be calculated in which positions the actuators should be to minimize the aberrations. The fine tuning is afterwards done via direct feedback from the AO. It turned out that the actuators for tilt had to be moved significantly in May 2019, which may support the hypothesis that remounting M1 introduces mirror tilt. In the future, we will explore making M1 actively controlled, which would enable us to perform this procedure in minutes instead of days.

2) Because M1 and M2 are not perfectly aligned with the optical axis, even via the above procedure, a beam wobble is introduced downstream. For the remaining small beam deviations, we programmed a ``beam tracker'', which consists of a webcam pointing to F3 and a program that automatically monitors the beam position and sends signals to tilt M5 to compensate for the diurnal variation. The beam tracker program fits a circle to the beam in F3 and keeps its position constant, which stabilized all focal planes after F3. Full stability can only be achieved if also the pupil is stabilized. This is done by measuring the intensity distribution on the WFS and an offset of the pupil is therefore directly visible. The AO controls the tilt of M11 and because the DM is exactly in a pupil plane, all subsequent pupils are stabilized. M11 is located 1000 mm from F3, while M5 is located 808 mm from a pupil. While both mirrors are not perfectly located in pupil or image planes, and therefore both influence the image and pupil motions, the effect on the pupil (when moving M5) or the image (when moving M11) planes is minimal and both trackers converge iteratively within seconds to a centered image in F3 and pupil on the DM.

\section{Infrastructure}\label{infrastructure}

\subsection{Vibrations}
Vibrations are a common problem in telescopes and depending on their frequencies cause significant issues in the data. We measured vibrations at GREGOR and improved several culprits that introduced vibrations: the mirror coolers were insulated better from the floor, an optical table was replaced with a more stable version, the adaptive optics setup was fortified with additional struts before it was made completely horizontal, additional supports were constructed for the slitjaw system, and the air conditioning is being investigated because we observed it to cause vibrations at 50, 60 and 125 Hz. We found that the air used for the cooling of M1, which is distributed to the backside of M1 via fans, seems to cause vibrations of M1 around 200 and 390 Hz.  Vibrations below 100 Hz can be mostly corrected by the AO system, but the peaks around 200 and 390 Hz unfortunately cannot and are also seen in AO power spectra. Their amplitude can be influenced by the fan speed of the M1 cooling and in certain outside temperature conditions, it is possible to reduce the amplitude by $\sim$ 75\%. However, a universally applicable solution is currently being investigated because a first attempt of stiffening the housing of the coolers did not reduce the vibrations. In general, the combined measures taken so far have reduced the overall vibrations by at least a factor of 10.

\subsection{Seeing}
As the air surrounding the telescope heats up during the day, the image quality degrades (``seeing''). At GREGOR this often translates into $r_0$ values that drop by a factor of 4 from early morning to afternoon. We therefore repainted all panels on the East side of the building with a special paint that keeps the panels within one degree of the ambient temperature by emitting in the infrared. Additionally, we repainted the flagstones and the tiles on the roof of the building with special paint based on titanium oxide and our temperature measurements with an infrared camera have shown that the floor temperature decreased by at least 2 degrees.

Temperature logging in the optics lab, which was installed in 4/2019, shows diurnal variations with an amplitude of about 0.5 $^\circ$C, which may contribute to the seeing in the optics lab and their cause is being investigated. As a first step, we created a new server room adjacent to the optics lab and are in the process of relocating most electronics/computers that contribute to heating the optics lab.

\subsection{Electric Installations}
In 2017 the old uninterruptible power supply (UPS) was replaced with a new 3-phased UPS, which doubled the power output (to 40 kVA). The new control of the UPS was programmed by KIS and installed by a company. Since 02/2019 all critical systems of GREGOR are on UPS. During power outages, the UPS bridges the time until the Diesel generator starts. In case of catastrophic failures, the UPS has enough power to supply GREGOR for about 30 minutes, which allows for a controlled shutdown, especially of the dome, which must be closed to protect the telescope from weather.

\subsection{Documentation}
Changes that are not documented can lead to frustration when systems behave differently than during the last time. We have therefore introduced a version-controlled repository of GREGOR documents, partially open to the public via \url{http://www.leibniz-kis.de/de/observatorien/gregor/documentation/}. These include manuals, checklists, technical notes, and requirements specifications. All instruments now have checklists that are also prominently displayed in the observing room, which simplifies the observations and leads to e.g. fewer missing calibration data. Additionally, every observer writes an observing report after their observing run detailing problems (if any) and every maintenance is documented with maintenance reports that are accessible to all partners via an internal website. Errors and bugs are reported via a ticketing system for better traceability of all actions.

\subsection{Observing room}
GREGOR's observing room is located on the third floor, two floors below the optics lab. It features three control computers, one of which is usually used for telescope and AO controls and the other two are available to control instruments. The screens can be mirrored in Freiburg for remote observations or observing support. We installed 4 webcams on top of the neighboring VTT telescope to monitor the clouds and a new weather screen is mounted in the observing room showing the 360$^\circ$ view, solar activity, and seeing and weather-plots. The background automatically turns red in case a weather limit is exceeded, for example if the wind exceeds 20 m/s or if the humidity exceeds 80\%, indicating the need to close the telescope. This has solved discussions between observers and assistants whether one should or should not close the telescope.


\section{Instrumentation}\label{control}
In this Section we only review upgrades made by KIS from 2017-2020. There were no changes to BBI, the GREGOR planet polarimeter (GPP) and the guest instrument ZIMPOL. 

\subsection{Slitjaw System}
The slitjaw (SJ) system is a context imager for the GRIS instrument. It was completely redesigned in 2018 to provide diffraction-limited images. The GRIS slit mask and integral field unit (IFU) mask are inclined by 15$^\circ$, which reflects the beam upwards (beam angle 2x15$^\circ$) as shown in the optical layout in Figure~\ref{sjoptics}. A motorized lens with a focal length of 750 mm automatically adjusts its position depending on the slit position during a scan to keep a constant distance between the (inclined) slit mask and the lens. The light is then distributed onto 3 horizontal rails above the SJ table. The lowest rail can be used for temporary setups. The middle rail gets 80\% of the incoming light and features an H-$\alpha$ Lyot filter with a passband of 0.4 \AA\ for chromospheric context images. The upper rail contains a 777 nm filter and shows photospheric context images. Both context images use a f=300 mm lens, which together with Prosilica GT2050 cameras with 2048 $\times$ 2048 pixels (5.5 $\mu$m $ \times$ 5.5 $\mu$m) lead to a plate scale of 0\farcs05 pixel$^{-1}$.

For stability reasons, the whole SJ/GRIS optical table was replaced, which reduced vibrations by one order of magnitude. Additionally, supports were constructed, which stabilize the overhanging part of the setup. A slit scanning device was developed, by building the mechanics and by programming the controls, which allows to scan areas of the Sun. 

  \begin{figure}[tb] %
  \centering 
   \includegraphics[width=0.45\textwidth]{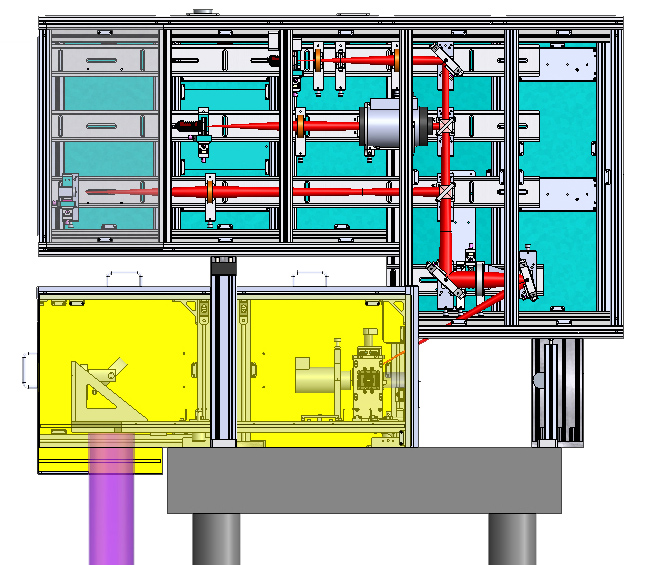}
   \caption{The SJ system installed at GREGOR. The upper two channels are continuum and H-alpha, respectively, while the bottom channel is currently not used.}
        \label{sjoptics}
  \end{figure}

\subsection{Adaptive Optics}
The numerous enhancements of the GREGOR adaptive optics  \citep{berkefeld2012AN} will be described in a forthcoming publication, with only a short summary given here. The main enhancements of the AO include that the AO takes control of the telescope in closed-loop operations. By sending small positioning commands to the telescope, based on the tip-tilt measurements, the AO can track the desired feature, thus compensating for solar rotation, or positioning issues, which may exceed the range of regular tip-tilt correction. It also controls the z-position of M2 for the optimal focus, which needs to be adapted during the day because of temperature and gravity changes, which lead to small variations in the focal distance between M1 and M2, but which are too large to be corrected by a simple focus offset via the DM. Furthermore, the AO automatically sets the ideal number of modes for the correction, depending on the seeing and the target, and can also update the reference image periodically if desired.


\section{Control Systems}\label{control}
 
Most of the GREGOR GUIs were designed to be engineering GUIs and therefore not particularly user-friendly for first-time observers. Before 2019, an observer therefore had to use more than 10 separate GUIs to control the telescope functions. A workshop was organized at KIS with representatives from each institute (KIS, MPS, AIP, IAC) during which requirements for a new observing GUI were defined that would simplify the operation and display all relevant functions, plus add some new functionality, in one GUI. The new GUI was developed within a year, and is operational since 2019.

 \begin{figure}[tb] 
  \centering 
   \includegraphics[width=0.49\textwidth]{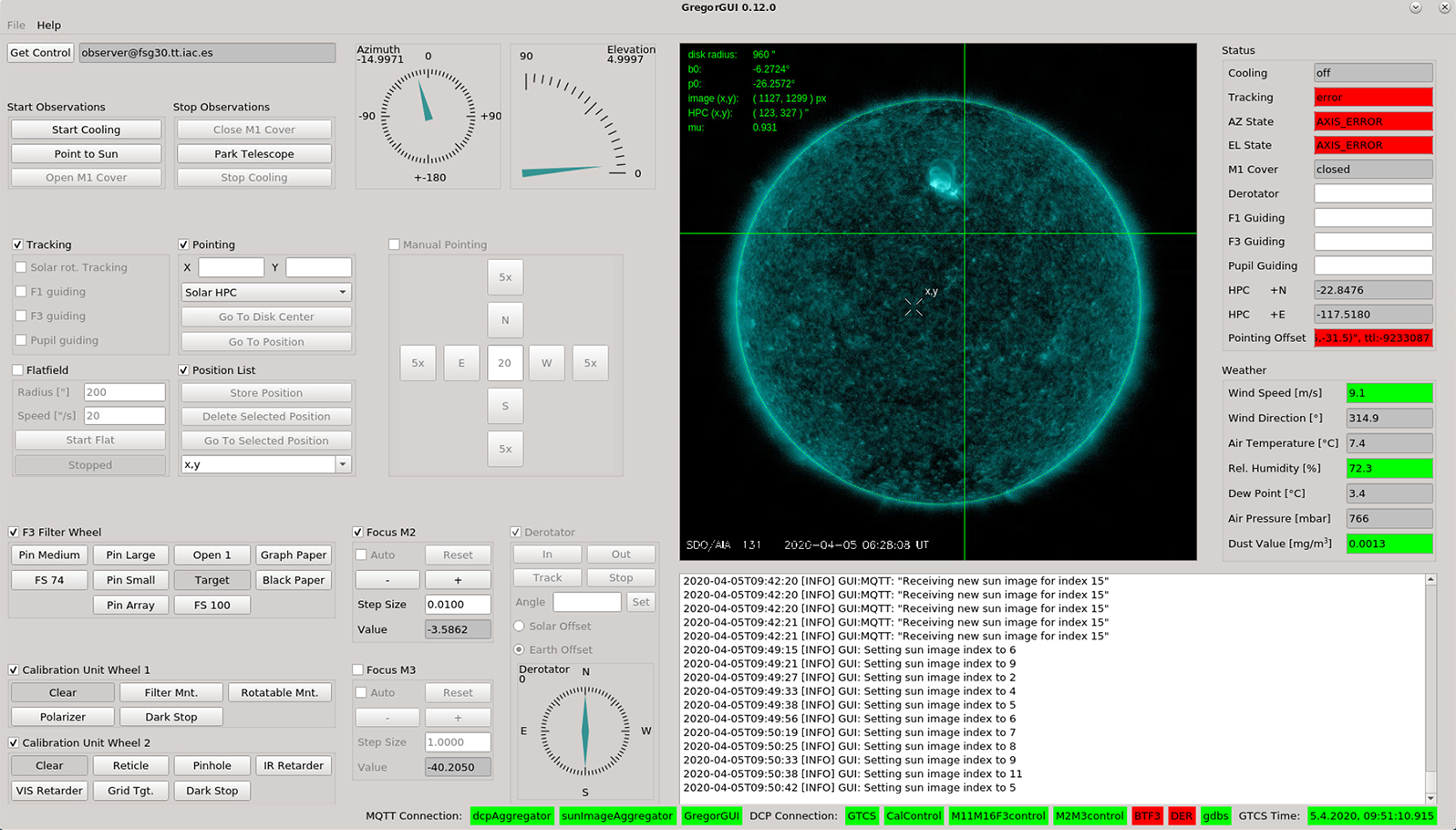}
   \caption{The GREGOR GUI. It allows us to control all essential telescope systems, such as positioning, pointing, filter wheels, calibration optics, focusing, and the derotator. The right side contains a display of current solar data, a system and weather status, and an event log.}
        \label{grggui}
  \end{figure}

{It is programmed in QT and based on a publisher-subscriber principle using MQTT as protocol. A data aggregation service collects all data from all subsystems and provides these data, plus external data such as current solar images published by other dedicated services, via topics on a MQTT broker. Multiple subscribers, e.g.\,GREGOR GUIs, can subscribe to these topics to be updated if their content changes.
The advantage of this approach is that future applications, for example a weather status page, can also be
programmed to subscribe to the MQTT broker to obtain the relevant data. This will avoid past
issues when too many simultaneous commands created issues with the network and system load. Safety-relevant features, such as who can take control, had to be considered
in addition to user-friendliness and simplicity of operation: User IDs are published using the same underlying mechanism creating a UID-specific command topic and the commander service subscribes to this to receive commands from only one commanding instance (e.g.\,a GREGOR GUI instance) at a time. }The layout of the GUI is shown in Fig.~\ref{grggui} and was structured, such that telescope operation is on the top, then pointing, positioning and flatfielding, and control of motors (filter wheels, derotator, focusing) on the bottom. The right side of the GUI displays the solar image in any desired SDO wavelength, status data (weather, systems, seeing) and a log file.

 %

 \section{Management}
 \subsection{Two observing seasons}
 In addition to the optical and facility upgrades, we also introduced policy changes. One was to make the observing time planning more flexible by having two observing seasons each year. Season 1 normally lasts from April-August and season 2 from August-December. This greatly facilitated dealing with unexpected issues, such as mirror recoatings that became urgent during the year. It also is advantageous for scientists when they do not have to wait for a year to apply for observing time if they have a good scientific idea right after a deadline.

\subsection{Access and permissions}
The other policy change was a much stricter control of telescope access and the limitation of personnel that is allowed to modify optics. Visitors no longer suddenly appear in the optics lab during observations. The alignment is also much more stable and there are fewer unexpected changes and issues. After 2 years of observing with the new system, the observing reports are generally very positive and campaigns without good data are rare.

\subsection{Technical Maintenance and Assistance}
GREGOR is a very complex telescope for its size. It is operated with a technical observing assistant on each observing day, who is responsible for the telescope operation and the hardware. About 200 person-days are spent for assistance, with assistant campaigns lasting 2-3 weeks. While this number is high, having qualified operators for the telescope and AO increased GREGOR's productivity significantly.

Technical maintenance is performed mostly during the winter break. More than 500 person-days are spent annually on technical work at the telescope, ranging from recoating mirrors, optics, mechanics, electronics, to updates of the infrastructure.


\section{Impact of the Upgrades on Observations}

Observers at GREGOR will notice many changes and enhancements. It is now possible to keep the AO locked on the quiet Sun for time periods exceeding a few dozen minutes. For objects with high contrast, such as sunspots and pores, the potential observing time spans hours. Observers will also notice a strong increase in image sharpness, especially outside of the central area of the field of view. This should enable studies of larger regions, for example active regions, including penumbrae, flows, large filaments, or flares. In terms of user-friendliness, the enhancements of the derotator, the control systems, especially the GREGOR GUI,  and the technical infrastructure, allow observers to quickly point the telescope to the desired position and image orientation, with fast changes possible for target of opportunity observations. Observers can also expect to be able to work more efficiently, because documentation and checklists are available for instruments and systems, which for example avoids that important calibration data are missed.

Several further enhancements are planned for future observations: A new spectropolarimeter based on Fabry Perot Interferometers is in development, which will allow quasi-simultaneous magnetic field measurements of a field of view exceeding 50\arcsec\ in the photosphere and in the chromosphere. Another planned upgrade involves the adaptive optics, which will be equipped with an H-$\alpha$ filter and thus be able to lock on e.g.\, prominences and spicules near the solar limb, thus improving the image stabilization and quality for off-limb observations.


\section{Summary and Conclusions}

GREGOR has undergone significant changes from 2018-2020. New relay optics allow us to reach diffraction-limited optical quality over the whole field of view. The new instrument distribution allows us to plan for future upgrades of instrumentation. The replacement of the last Cesic mirror with one made of Zerodur improved the contrast noticeably. A new slitjaw system provides diffraction-limited context images in the photosphere and in the chromosphere. The adaptive optics was enhanced to optimize many of the previously manual telescope functions, such as tracking and focusing. The derotator functionality was enhanced and its orientation calibrated.

New alignment methods were developed and solutions were found to optimize the beam stability and reduce the vibrations. The telescope is now controlled by a dedicated GUI, which includes the functionality for all subsystems. The observing room was optimized, with dedicated work stations to control the telescope and instruments and a weather monitor to improve the safety of noticing weather-related telescope closure conditions. Most systems are documented with manuals, checklists and technical notes, which are accessible in a version-controlled database,  based on a model for documentation of space missions. Management and policy changes further improved the stability of the telescope and consistently obtaining scientifically valuable data.

\begin{acknowledgements}
\textit{Operating GREGOR would not be possible without the dedicated KIS technical staff who are all very gratefully acknowledged. In particular, we would like to thank Oliver Wiloth, Andreas Bernert, Stefan Semeraro, Frank Heidecke, Michael Weisssch\"adel, Clemens Halbgewachs, Tobias Preis, Peter Markus, Peter Caligari, Marco G\"unter, Markus Knobloch, Roland Fellmann, Christine Fellmann, Bruno Femenia, Daniel Gisler, and Sylvia Nadler. We also thank G\"oran Scharmer and an anonymous referee for helpful comments on the manuscript.
The 1.5-meter GREGOR solar telescope was built by a German consortium under the leadership of the Leibniz-Institute for Solar Physics (KIS) in Freiburg with the Leibniz Institute for Astrophysics Potsdam, the Institute for Astrophysics G\"ottingen, and the Max Planck Institute for Solar System Research in G\"ottingen as partners, and with contributions by the Instituto de Astrofisica de Canarias and the Astronomical Institute of the Academy of Sciences of the Czech Republic. }
\end{acknowledgements}

\bibliographystyle{apj}
\bibliography{journals,ibisflare}

\end{document}